\documentclass[prd,a4paper,showpacs,preprintnumbers,nofootinbib]{revtex4} 

\usepackage{epsfig}

\newcommand{\bm}[1]{\mbox{\boldmath $#1$}}

\newcommand{\open}{{<\kern -0.3 em{\scriptscriptstyle )}}}

\newcommand{\nslash}{\kern 0.2 em n\kern -0.45em /}
\newcommand{\Pslash}{\kern 0.2 em P\kern -0.56em \raisebox{0.3ex}{/}}
\newcommand{\pslash}{\kern 0.2 em p\kern -0.4em /}
\newcommand{\kslash}{\kern 0.2 em k\kern -0.45em /}
\newcommand{\Sslash}{\kern 0.2 em S\kern -0.56em \raisebox{0.3ex}{/}}

\newcommand{\eq}{\begin{equation}}
\newcommand{\ee}{\end{equation}}
\newcommand{\beq}{\begin{equation}}
\newcommand{\eeq}{\end{equation}}
\newcommand{\ba}{\begin{eqnarray}}
\newcommand{\ea}{\end{eqnarray}}
\newcommand{\eqa}{\begin{eqnarray}}
\newcommand{\eea}{\end{eqnarray}}

\newcommand{\sumint}{\kern 0.2 em {\textstyle\sum} \kern -1.1 em \int}

\newcommand{\la}{\langle}
\newcommand{\ra}{\rangle}
\newcommand{\amp}[1]{\la #1 \ra}

\begin{document} 

\title{Asymmetric jet correlations in $p \, p^\uparrow$ scattering}

\author{Dani\"el Boer}
\email{dboer@nat.vu.nl}
\affiliation{Dept.\ of Physics and Astronomy, Vrije Universiteit Amsterdam, \\
De Boelelaan 1081, 1081 HV Amsterdam, The Netherlands}

\author{Werner Vogelsang}
\email{vogelsan@quark.phy.bnl.gov}
\affiliation{Physics Department and RIKEN-BNL Research Center\\ 
Brookhaven National Laboratory, Upton, NY 11973, U.S.A.}

\date{\today}

\begin{abstract}
We propose that back-to-back correlations in azimuthal angle of 
jets produced in collisions of unpolarized with transversely polarized 
proton beams could be used to determine Sivers functions. 
The corresponding single-spin asymmetry is not power-suppressed, but is
subject to Sudakov suppression. We present estimates of the 
asymmetry (without and with Sudakov effects) for RHIC at
jet transverse momenta of $\sim 10$~GeV and
show that it may reach a few per cent or more and could provide
access to the gluon Sivers function. 
\end{abstract}

\preprint{BNL-NT-03/42, RBRC-390}

\maketitle


\section{Introduction}

An important goal of ongoing experiments with polarized protons
at BNL's Relativistic Heavy-Ion Collider RHIC is to contribute
to a better understanding of transverse-spin effects in QCD.
Of particular interest are single-spin asymmetries $A_{\mathrm{N}}$, 
obtained from scattering one transversely polarized proton off an 
unpolarized one. It was found a long time ago in fixed-target
experiments \cite{E704,AGS} that such $p^{\uparrow}p$ collisions 
can yield strongly asymmetric distributions of hadrons in the final 
state. The most famous examples are the sizable (${\cal O}(10\%)$)
$A_{\mathrm{N}}$ found in the process $p \, p^\uparrow \to \pi \, X$, 
which express the fact that the produced pions have a preference to 
go to a particular side of the plane spanned by the proton beam 
direction and the transverse spin direction. Recently, the 
STAR collaboration at RHIC has found that such large asymmetries
persist even at collider energies \cite{STAR03}. It is fair to say that 
to date the asymmetries have defied a full understanding at the 
quark-gluon level in QCD. One reason for this is that in QCD 
$A_{\mathrm{N}}$ for a single-inclusive reaction is power-suppressed 
as $1/p^{\perp}$ in the hard scale given by the transverse momentum 
$p^{\perp}$ of the pion. This makes the formalism for describing
the asymmetries rather complicated, compared to more standard 
leading-power observables in perturbative QCD. An attempt of an
explanation for the observed $A_{\mathrm{N}}$ has been given within 
a formalism \cite{qs} that systematically treats the 
power-suppression of $A_{\mathrm{N}}$ in terms of higher-twist parton 
correlation functions.  Alternatively, it has been 
proposed \cite{Sivers,Collins-93b,Anselmino,Boer-00} 
that the dependence of parton distributions and fragmentations 
functions on a small ``intrinsic'' transverse momentum 
$k^{\perp}$ could be responsible for the asymmetries, 
through the interplay with the partonic elementary cross sections
that are functions steeply falling with $p^{\perp}$. $A_{\mathrm{N}}$ 
is generated from the $k^{\perp}$-odd parts of the partonic scatterings, 
which acquire an additional factor $1/p^{\perp}$, making the
mechanism again effectively higher twist.
Measurements of just $A_{\mathrm{N}}$ in $p \, p^\uparrow \to \pi \, X$
will not be sufficient to disentangle all these effects, and it has
also been shown recently that for this power-suppressed observable
the mechanisms are not all independent of one another \cite{BMP03}.

There is, however, a class of observables for which the
$k^{\perp}$-dependent distributions or fragmentation functions alone
are relevant, and may actually lead to leading-power effects. These 
are observables directly
sensitive to a small measured transverse momentum. In spin physics,
the most well-known example in this class was given by Collins 
\cite{Collins-93b}. He proposed to consider the single-transverse
spin asymmetry in semi-inclusive deeply-inelastic scattering (SIDIS),
$e\, \vec{p}\to e' \, \pi \, X$, where the pion is detected out of the 
scattering plane. This asymmetry may receive contributions from the
$k^{\perp}$ effects mentioned above: from the
transverse momentum of the pion relative to its quark 
progenitor (the so-called Collins effect \cite{Collins-93b}), or from the
intrinsic $k^{\perp}$ of a parton in the initial proton
(referred to as Sivers mechanism \cite{Sivers}). Here we propose a new
observable sensitive to the latter effect. 

More precisely, the Sivers effect is a correlation between the 
direction of the transverse spin of the proton and the transverse 
momentum direction of an unpolarized parton inside the proton \cite{Sivers}. 
The Sivers effect in the process $p \, p^\uparrow \to \pi \, X$ has first 
been analyzed in detail 
by Anselmino {\it et al.}~\cite{Anselmino}, who extracted the 
Sivers functions for valence quarks from a fit to the data under the assumption
that the asymmetry is solely due to this effect. Subsequently, the single spin
asymmetry for the Drell-Yan process, which is another 
process that belongs to the 
class of ``leading-power'' observables mentioned above, was predicted 
\cite{ADM}. 

Following in part the notation of Ref.\ \cite{ADM}, the number density 
of a parton $f=u,\bar{u},\ldots,g$ inside a proton with transverse 
polarization ${\bm S}_T$ and three-momentum ${\bm P}$, is parameterized as
\beq
\hat f (x,{\bm k}^{\perp},{\bm S}_T) = f (x,k^{\perp}) + 
\frac{1}{2} \Delta^N f (x,k^{\perp}) \frac{{\bm S}_T \cdot
({\bm P} \times {\bm k}^{\perp})}{|{\bm S}_T| \, |{\bm P}| \, 
|{\bm k}^{\perp}|},
\label{SiversADM}
\eeq
where ${\bm k}^{\perp}$ is the quark's transverse momentum, with
${k}^\perp = | {\bm k}^\perp |$. The function $f (x,k^{\perp})$ is the 
unpolarized parton distribution, and $\Delta^N f$ denotes the Sivers
function. As one can see, the correlation proposed by Sivers
corresponds to a time-reversal odd triple product ${\bm S}_T \cdot
({\bm P} \times {\bm k}^{\perp})$. Since for stable initial
hadrons no strong interaction phases are expected, until recently it was
widely believed that the Sivers functions
had to vanish identically. It was then discovered
\cite{BHS,Belitsky,Collins-02,BMP03}, however, that the 
time-reversal symmetry argument against the Sivers functions is invalidated 
by the presence of the Wilson lines in the operators defining the 
parton densities. These are required by gauge invariance and, as
under time reversal future-pointing Wilson lines turn into past-pointing 
ones, the time reversal properties of the Sivers functions are non-trivial
and permit them to be non-vanishing.
It is intriguing that the possibility of a non-vanishing
Sivers function emerges solely from the Wilson lines in QCD.  
Another aspect to the physics importance of the Sivers function is the 
fact that it arises as an interference of wave functions 
with angular momenta $J_z=\pm 1/2$ and hence contains information
on parton orbital angular momentum \cite{BHS,Ji,Burkardt}. 

In this paper we will discuss a single-spin asymmetry in $p\,p$ 
scattering that also belongs to the class of ``leading-power'' observables. 
The reaction we will consider is the inclusive 
production of jet pairs, $p\, p^\uparrow \to\mathrm{jet}_1 \, 
\mathrm{jet}_2 \, X$, for which the two jets are nearly back-to-back 
in azimuthal angle. 
This requirement makes the jet pairs sensitive to a 
small measured transverse momentum, and hence allows the single-spin
asymmetry for the process to be of leading power. The asymmetry 
$A_{\mathrm{N}}$ for this 
process should give direct access to the Sivers function. Actually,
in contrast to SIDIS and to $p^{\uparrow}p\to \pi X$, the observable we
propose here has the feature that it is rather sensitive to the 
nonvalence contributions to the Sivers effect, in particular
the gluon Sivers function. The latter has not been considered so far
in any phenomenological asymmetry study. Its precise definition has been
given by Mulders and Rodrigues \cite{Joao}. They define a gluon 
correlation function (in the light-cone ($A^+=0$) gauge)
\beq
M \Gamma^{ij}(x,{\bm k}_T) \equiv \int \frac{d \xi^-  d^2 \xi_T}{(2\pi)^3}
e^{i k\cdot \xi} \amp{P,S | F^{+i}(0) F^{+j}(\xi)|P,S}|_{\xi^+=0}\;,
\label{gluonDF}
\eeq
which can be parameterized in terms of gluon distribution functions. In
particular, one has
\beq
M \Gamma_{ij}(x,{\bm k}_T) g_{T}^{ij} = - x P^+ \left[ G(x,{\bm k}_T^2) +
\frac{\epsilon_T^{k_T S_T}}{M} G_T(x,{\bm k}_T^2) \right]\; ,
\label{gluonSivers}
\eeq
where $ G(x,{\bm k}_T^2)$ is the unpolarized transverse momentum 
dependent gluon distribution inside an unpolarized hadron.
Eqs.\ (\ref{gluonDF}) and (\ref{gluonSivers}) can be viewed as an explicit 
definition of Eq.\ (\ref{SiversADM}) in the gluon sector, i.e., $G_T$ 
corresponds to $\Delta^N g(x,k^{\perp})$, up to a 
$k^{\perp}$ dependent normalization factor.

As the above shows, the inclusion of transverse momentum dependent parton
distributions is necessary when discussing Sivers effect asymmetries. This 
extension of the ordinary parton distributions that are functions of the
lightcone momentum fractions only is not straightforward from a theoretical
point of view. Factorization theorems involving $k^{\perp}$ dependence
are generally harder to derive. Also, Sudakov suppression effects 
become relevant. As is well-known from unpolarized hadron-hadron 
collisions, the average transverse momenta of pair final states,
naively associated with originating from intrinsic transverse
momentum, are energy ($\sqrt{s}$) dependent and can reach several 
GeV at collider energies \cite{begel,apan1}. Clearly, such large 
average transverse momenta are not to be attributed to intrinsic 
transverse momenta alone, but mostly to the transverse momentum broadening
due to (soft) gluon emissions. In our study we will therefore include
such Sudakov effects, albeit within a somewhat simplified treatment. 
We will use experimental data to obtain estimates for 
the average transverse momenta of initial and radiated partons in unpolarized 
hadron-hadron collisions, which we will subsequently use to obtain 
estimates for the Sivers effect asymmetry in the azimuthal angular 
distribution of jets with respect to opposite side jets.
We will do this without and with inclusion of Sudakov factors; 
one has to keep in mind that in the first case the average ``intrinsic'' 
transverse momenta we will find will effectively contain a significant 
perturbative (Sudakov) component. It will be instructive to compare 
the results of the two analyses. 

Our study is also motivated by the favorable experimental situation 
at RHIC, where the STAR collaboration has recently presented data
\cite{STAR} for a closely related back-to-back reaction. 
In the next section we will discuss back-to-back correlations
in the unpolarized case and compare to the STAR data. 
In Sec.~III we will then address the spin asymmetry in
$p\, p^\uparrow \to\mathrm{jet}_1 \, \mathrm{jet}_2 \, X$.
Section~IV presents our conclusions and a further discussion of
some theoretical issues.

\section{Jet correlations in unpolarized hadron collisions}

We first consider the inclusive production of jet pairs
in unpolarized proton-proton collisions, $p\, p\to\mathrm{jet}_1 \, 
\mathrm{jet}_2 \, X$. We take each of the jets to have large
transverse momentum. This implies the presence of 
short-distance phenomena, which may be separated
from long-distance ones. More precisely, the cross
section for this process factorizes into convolutions
of parton distribution functions with partonic hard-scattering
cross sections that may be evaluated using QCD perturbation
theory.  To lowest order, the partonic subprocesses are the
QCD two-parton to two-parton scatterings. If the cross
section observable is defined in such a way that it is
insensitive to transverse momenta of the initial partons or
to particles radiated at small transverse momentum,
the factorization is the standard ``collinear'' one, 
and the convolutions are simply in terms of (light-cone) momentum 
fractions. 
The observable we are interested in is slightly more involved. 
We choose the two jets to be almost back-to-back 
when projected into the plane perpendicular to the direction of the beams, 
which is equivalent to the jets being separated by nearly
$\Delta\phi\equiv\phi_{j_2}-\phi_{j_1}=\pi$ in azimuth. Such 
a configuration directly corresponds to an observed small
transverse momentum of the jet pair. 
In the case of two-by-two scattering of collinear initial partons, 
the jets are exactly back-to-back in azimuth. Deviations from this may result 
from additional partons being radiated into the final state. If 
the jets have fairly large separation in azimuthal angle, the 
dominant contribution to the cross section will come from a 
single additional parton radiated into the final-state against 
which the two jets recoil. Closer to $\Delta \phi=\pi$, radiation 
is suppressed, and Sudakov effects become relevant\footnote{
Hard three (or more) jet 
configurations may also contribute near $\Delta \phi=\pi$,
if the additional hard parton happens to be almost in one
plane with the two jets. Such situations are expected to be 
relatively rare because they are not associated with singular 
behavior of the perturbative cross section in the azimuthal 
back-to-back region. We ignore them in our analysis.}. Intrinsic 
transverse momenta of the initial partons may become important
as well. 

In such nearly back-to-back situations, factorization is not necessarily 
lost; rather, a factorization theorem now needs to be formulated in terms
of parton distributions depending on light-cone momentum
fraction {\em and\/} transverse momentum. Factorization theorems of this type
have been discussed for the simpler process $e^+ \, e^- \to A \, B \, X$, 
where $A$ and $B$ are two hadrons almost back-to-back \cite{CS81,CS85e}, and 
for Drell-Yan type processes \cite{LSV}. For factorization to occur it is 
essential that transverse momenta of initial or radiated partons are linked 
to the observed small transverse momentum only kinematically,
that is, by momentum conservation, but are neglected in the
hard scattering. For $p\, p\to\mathrm{jet}_1 \, 
\mathrm{jet}_2 \, X$, factorization at small measured transverse 
momentum of the pair has to our knowledge not yet been proven explicitly, 
but here we will assume that it falls in the class discussed in Ref.\ 
\cite{LSV}.

In the following we first consider only intrinsic transverse
momentum of the initial partons and neglect perturbative 
radiation of particles into the final state and Sudakov effects. 
The latter will generally be very relevant and we will include them 
afterwards to leading logarithm for the purpose of estimating them.
A treatment beyond leading logarithm is fairly complicated for $p\, 
p\to\mathrm{jet}_1\, \mathrm{jet}_2 \, X$ and not within the scope of 
this work. Our treatment follows the same strategy as 
applied in Refs.\ \cite{Boer-00,Boer-01}. 

In the absence of additional partons being radiated into the 
final state, momentum conservation implies that the sum of the 
transverse momenta of the two jets is equal to the sum of the transverse 
momenta of the initial state partons. To be more explicit, we expect 
the cross section to be proportional to 
\beq \label{denom0}
{\mathcal U}=\int d^2k^{\perp}_1 d^2k^{\perp}_2 \; \delta^2 
\left( {\bm k}_1^{\perp} + {\bm k}_2^{\perp} - 
{\bm P}_{j_1}^{\perp}-{\bm P}_{j_2}^{\perp}\right) \;
f_1(k^{\perp}_1) \;f_2(k^{\perp}_2)
\; ,
\eeq
where ${\bm P}_{j_1}^{\perp}$ and ${\bm P}_{j_2}^{\perp}$ are 
the transverse momenta of the jets, and the $f_i$ are the 
transverse momentum distributions of the initial partons. In general, 
the $f_i$ will also depend on the parton lightcone momentum
fractions. The factor ${\mathcal U}$ may be thought of
as a smeared-out $\delta^2\left( {\bm P}_{j_1}^{\perp}+
{\bm P}_{j_2}^{\perp} \right)$ representing the standard 
transverse-momentum conservation for collinear partons.

Since the distributions $f_i$ are not known, we will assume
Gaussian transverse momentum dependence for simplicity:
\beq \label{gauss}
f_i(k^{\perp}_i)=\frac{\mathrm{e}^{-\left(k^{\perp}_i\right)^2/
\langle k^{\perp\,2}_i\rangle}}{\pi\langle k^{\perp\,2}_i\rangle} \; .
\eeq
Moreover, we will assume that the average transverse momentum squared is the
same for all partons in the proton and independent of $x$, 
i.e.\ $\langle k^{\perp\,2}_i\rangle  
\equiv \langle k^{\perp\,2}\rangle$ for $i=1,2$. 
These simplifications may all be
improved upon at a later stage, when there are data requiring a more 
sophisticated treatment, but here we will focus on the proof of 
principle rather than on making an accurate quantitative prediction. 
The assumption of Gaussians is for convenience and sufficient for our 
purpose. One obtains 
\beq
{\mathcal U}=\frac{\mathrm{e}^{-(r^{\perp})^2/\left(
\langle k^{\perp\,2}_1\rangle+\langle k^{\perp\,2}_2\rangle\right)}}
{\pi\left(\langle k^{\perp\,2}_1\rangle+\langle k^{\perp\,2}_2\rangle
\right)} \;  ,
\label{denom}
\eeq
where ${\bm r}^{\perp}={\bm P}_{j_1}^{\perp}+{\bm P}_{j_2}^{\perp}$. One has 
\beq
({\bm r}^{\perp})^2 = P_{j_1}^{\perp\,2}+P_{j_2}^{\perp\,2}+
2 P_{j_1}^{\perp} P_{j_2}^{\perp} \cos (\Delta \phi) \; ,
\eeq
where $\Delta \phi$ is the separation of the two jets in azimuth. 
Hence, one finds that 
\beq \label{denom1}
{\mathcal U} \propto e^{-K_{{\mathcal U}} \cos(\Delta \phi)},
\eeq
where $K_{{\mathcal U}}=P_{j_1}^{\perp} P_{j_2}^{\perp} /
\langle k^{\perp\,2}\rangle$. This implies
that ${\mathcal U}$ is peaked around $\Delta \phi = \pi$ as expected.
Expansion for small $\delta \phi\equiv \Delta \phi - \pi$ shows
a Gaussian behavior near the peak. 

\begin{figure}[t]
\begin{center}
\epsfig{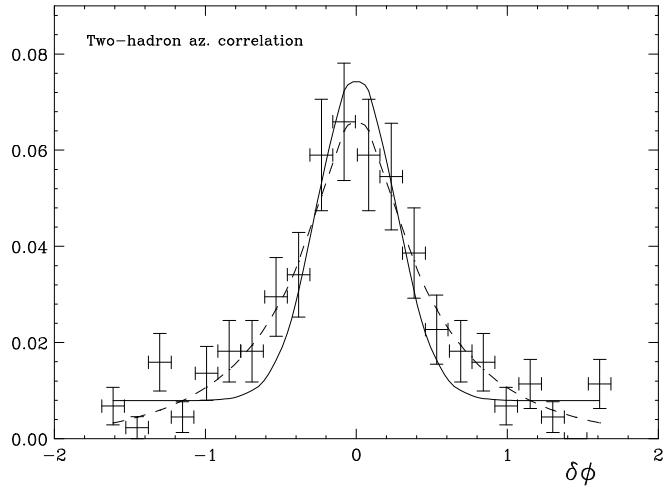}
\end{center}
\vspace*{-0.5cm}
\caption{Two-hadron azimuthal correlation in the back-to-back 
region. See Ref.\ \cite{STAR} for details on the experimental 
definition of the correlation function.
The data are from \cite{STAR}, and the solid curve is
obtained using the distribution (\ref{denom1}) and the
widths in Eq.~(\ref{width}), fitting the overall normalization
and an additive constant. The dashed line shows the 
result of a fit to the data with inclusion of leading-logarithmic
Sudakov effects. 
\label{fig:fig1}} 
\vspace*{-0.5cm}
\end{figure}

Correlations in $\Delta \phi$ for dijets have
been studied in \cite{isr}. Measurements of $\Delta \phi$ distributions 
have also been performed both at the ISR \cite{isrhad}, and in the 
fixed-target experiment E706 \cite{begel,apan1}, albeit not for two-jet 
correlations, but rather for pairs of nearly back-to-back leading hadrons,
usually pions. 
Recently, the STAR collaboration \cite{STAR} at RHIC (BNL) has presented 
precise data on hadron $\Delta \phi$ correlations from unpolarized 
proton-proton collisions at $\sqrt{S}=200$~GeV (and from heavy ion 
collisions). Our study of spin effects in back-to-back reactions in the 
next section will be tailored to $p\,p^{\uparrow}$ collisions at 
$\sqrt{S}=200$~GeV at RHIC, so we will compare our approach
to the STAR $\Delta \phi$ distribution data \cite{STAR}, displayed
for ``same-sign'' hadrons in $-\pi/2\leq \delta \phi \leq \pi/2$
in Fig.~\ref{fig:fig1}. In the STAR analysis, 
the first (``trigger'') hadron was required to
have 4~GeV~$\leq P^{\perp}_{h_1}\leq$~6~GeV, and the recoiling hadron
2~GeV~$\leq P^{\perp}_{h_2}\leq P^{\perp}_{h_1}$. The pseudorapidities
of both hadrons were within $|\eta_{h_{1,2}}|\leq 0.7$. As can be seen
from Fig.~\ref{fig:fig1}, the peak at $\delta \phi=\Delta \phi-\pi=0$ 
is clearly pronounced and appears consistent with a Gaussian behavior 
in $\delta \phi$. One also notices that the distribution does not decrease 
to zero at large $\delta \phi$, which could be indicative
of the perturbative tail corresponding to hard-gluon emission. 

The above considerations for dijet production can be modified 
to apply to dihadron production. The first modification is 
that one needs to take into account fragmentation functions
describing the probability with which a final-state
parton emerging from the hard scattering will yield
the observed hadron. The hadron will take a light-cone
momentum fraction $z$ of the parton momentum $p$, i.e.\
in particular for a collinear fragmentation process:
\beq
{\bm P}^{\perp}_{h_1}=z_1 {\bm p}^{\perp}_1 \; , \;\;\;\;
{\bm P}^{\perp}_{h_2}= z_2 {\bm p}^{\perp}_2 
\eeq
for the transverse components, and hence ${\bm r}^{\perp}=
{\bm P}_{h_1}^{\perp}/z_1+{\bm P}_{h_2}^{\perp}/z_2$ in the
formulas above. For our estimate
we will use in the following only average values for the
hadron transverse momenta over the experimental bins:
$\langle P^{\perp}_{h_1}\rangle \approx 4.5$~GeV, $\langle 
P^{\perp}_{h_2}\rangle\approx 2.5$~GeV.
We have then determined the corresponding average $z_i$ in
the framework of a leading-order calculation of the 
unpolarized dihadron cross section in $p\,p$ scattering at 
RHIC energy $\sqrt{S}=200$~GeV and find $\langle z_1 \rangle 
\approx 0.45$, $\langle z_2 \rangle \approx 0.25$. Here
we have used the CTEQ-5 \cite{cteq5} set of parton distribution 
functions and the fragmentation functions of \cite{kkp}, both at 
leading order.

Additionally, there will be a transverse-momentum smearing also
in the final state, that is, the observed hadron may be produced
at some small transverse momentum relative to the parent 
parton \cite{bep}, implying
\beq
{\bm P}^{\perp}_{h_1}= z_1 {\bm p}^{\perp}_1 + 
{\bm{\hat{k}}}^{\perp}_1 \; , \;\;\;\;
{\bm P}^{\perp}_{h_2}= z_2 {\bm p}^{\perp}_2  + 
{\bm{\hat{k}}}^{\perp}_2 \; ,
\eeq
with ${\bm p}^{\perp}_i\cdot {\bm{\hat{k}}}^{\perp}_i=0$. 
This will also have an influence on the $\delta \phi$ distribution 
of the two produced hadrons. We estimate this effect
by replacing $\langle k^{\perp\,2}_1\rangle+\langle k^{\perp\,2}_2
\rangle$ in Eq.~(\ref{denom1}) by 
$\langle k^{\perp\,2}_1\rangle+\langle k^{\perp\,2}_2\rangle+
\langle \hat{k}^{\perp\,2}_1\rangle/\langle z_1\rangle^2
+\langle \hat{k}^{\perp\,2}_2\rangle/\langle z_2\rangle^2$,
where $\langle \hat{k}^{\perp\,2}_i\rangle$ is the
average transverse momentum broadening squared in 
fragmentation. We could improve this estimate by
taking into account that for a final-state particle
emitted at some angle only a certain projection
of ${\bm{\hat{k}}}^{\perp}$ is relevant for the $\delta \phi$
distribution. This modifies the functional form 
of the distribution in $\delta \phi$. However, we found this 
effect to be rather insignificant for the $\delta \phi$ and
the widths we consider below. 

Our next goal is to obtain an estimate of the average $\langle 
k^{\perp\,2}_i\rangle$ at RHIC energy $\sqrt{S}=200$~GeV 
from a comparison to experimental data. For this, Ref.\ \cite{apan} 
is particularly useful, where an analysis of a variety of 
data from fixed-target and collider experiments
on transverse momenta of dimuon, diphoton, and dijet (or dihadron) 
pairs was performed. If one neglects radiative effects, 
such pair transverse momenta are directly related to intrinsic 
transverse momenta. The results of \cite{apan} show that the
pair transverse momenta increase with center-of-mass energy.
Also, they are consistently larger for dihadron pairs than
for diphoton or dimuon pairs. This may be understood from 
the presence of $k^{\perp}$ smearing in fragmentation. 
Additionally, in contrast to diphotons or dimuons, dihadron cross 
sections are dominated by scatterings of initial gluons, which
may have a somewhat larger $k^{\perp}$ broadening. As mentioned 
above, we neglect this effect. From the results shown in \cite{apan} 
we then estimate for dihadrons at $\sqrt{S}=200$~GeV
\beq \label{rel}
\langle k^{\perp\,2}_1\rangle+\langle k^{\perp\,2}_2\rangle+
\langle \hat{k}^{\perp\,2}_1\rangle/\langle z_1\rangle^2
+\langle \hat{k}^{\perp\,2}_2\rangle/\langle z_2\rangle^2\;
\approx\; 15 \;{\mathrm{GeV}}^2 \; ,
\eeq
and 
for non-fragmentation final states
\beq \label{rel1}
\langle k^{\perp\,2}_1\rangle+\langle k^{\perp\,2}_2\rangle\;
\approx\; 9 \;{\mathrm{GeV}}^2 \; .
\eeq
From this, we estimate, assuming $\langle k^{\perp\,2}_1\rangle=
\langle k^{\perp\,2}_2\rangle$ and $\langle \hat{k}^{\perp\,2}_1\rangle=
\langle \hat{k}^{\perp\,2}_2\rangle$:
\beq \label{width}
\sqrt{\langle k^{\perp\,2}_i\rangle}\approx 2\;\mathrm{GeV}\;,
\;\;\;\;\;\; \sqrt{\langle \hat{k}^{\perp\,2}_i\rangle}\approx 
0.5\;\mathrm{GeV} \; .
\eeq
Our value for the width for the initial-state broadening is larger
than that found from studies of single-particle inclusive
cross sections at the lower fixed-target energies in \cite{ADM}.
The curve in Fig.~\ref{fig:fig1} shows our result for the $\delta
\phi$ distribution based on the widths in Eq.~(\ref{width}).
We have fitted the overall normalization of the curve to the
data, and we have also allowed an additive constant to the
$\delta \phi$ distribution in this fit, in order to account
for the perturbative tail at larger $|\delta \phi|$. 
The resulting curve gives a fair description of the data,
even though the data appear to prefer a somewhat larger
width of the peak. 

The fairly large size of $\sqrt{\langle k^{\perp\,2}_i\rangle}$
in Eq.\ (\ref{width}) (as compared to typical hadronic mass scales) 
again indicates that there are significant perturbative effects 
that should be taken into account in a more thorough analysis. 
As we mentioned earlier, Sudakov effects, related to multi-soft-gluon
emission, are expected to be particularly relevant. Near 
$\delta \phi=0$, gluon radiation is kinematically inhibited, 
and the standard cancelations of infrared singularities between 
virtual and real diagrams lead to large logarithmic remainders in the
partonic hard-scattering cross sections. For the $\delta \phi$
distribution, these have the form $\alpha_s^k \ln^{2k-m}
(\delta \phi)/\delta \phi$ in $k$th order of perturbation theory,
with $1\leq m \leq 2k$ or, more generally, for the ${\bm r}^{\perp}$
distribution they are of the form $\alpha_s^k \ln^{2k-m}
(\hat{s}/|{\bm r}^{\perp}|^2)/|{\bm r}^{\perp}|^2$ \cite{PP,CSS},
where $\sqrt{\hat{s}}$ is the partonic center-of-mass energy.
It is possible to resum these logarithmic contributions to all orders 
in $\alpha_s$. For the leading logarithms ($m=1$), 
this was achieved a long time ago \cite{PP,DDT,BN}.
Recent progress in the resummation for $p\, p\to A\, B \, X$ 
at next-to-leading logarithmic level was reported in \cite{guffanti}. 
Here we provide an estimate of the Sudakov effects by
taking into account the tower of leading logarithms. 

Applying the derivation of \cite{PP} to the process $p\, 
p\to\mathrm{jet}_1\, \mathrm{jet}_2 \, X$, the resummation of independent 
soft gluon emissions to leading double logarithmic order leads to the 
following distribution in ${\bm r}^\perp$:
\beq \label{suda}
{\mathcal U}^{\mathrm{\,Sud}}\equiv
\frac{1}{\sigma_0}\frac{d\sigma}{d^2 r^\perp} = \int_0^\infty 
\frac{d b^2}{4\pi} \, 
J_0(b r^\perp) \, \exp \left[ -\frac{\alpha_s}{\pi} \,C\,\ln^2 
(b \sqrt{\hat{s}}) \right] \, \tilde{f}_1(b^2)\tilde{f}_2(b^2),
\eeq
where $\sigma_0$ is the lowest-order cross section integrated over 
all $r^\perp$, and $C$ is the sum of the color charges for the external
legs in the partonic hard scattering, i.e., for subprocesses
involving quarks and antiquarks only one has $C=4C_F=16/3$,
for processes with two quarks and two gluons $C=2(C_A + C_F) = 26/3$,
and for $gg \to gg$, $C=4C_A=12$. All these channels are relevant, 
because of the competition between the magnitude of the parton 
distributions and the magnitude of the Sudakov suppression. 
Finally, in Eq.~(\ref{suda})
\beq
\tilde{f}_i(b^2) = \int d^2 k_i^\perp \, e^{i {\bm b}\cdot {\bm k}_i^\perp}
f_i(k_i^\perp) = e^{-b^2 \langle k^{\perp\,2}_i\rangle/4} \; ,
\eeq
where the last equality follows for our Gaussian 
$k^\perp$ distributions of Eq.~(\ref{gauss}). In lowest 
order in $\alpha_s$, i.e.\ setting $\alpha_s=0$ in the Sudakov
exponent, Eq.~(\ref{suda}) reduces to ${\mathcal U}$ of Eq.\ (\ref{denom0}).

We have used the Sudakov improved ${\mathcal U}^{\mathrm{\,Sud}}$ 
of Eq.\ (\ref{suda}) in a fit of the value for 
$\langle k^{\perp\,2}\rangle$ to the STAR data. 
We find that inclusion of the leading logarithms 
leads to a markedly better agreement with the data,
demonstrated by the dashed line in Fig.~\ref{fig:fig1},
and to a reasonably small value of $\sqrt{\langle k^{\perp\,2}\rangle}
\approx 0.9$~GeV, which is closer to a typical hadronic mass scale and to 
the one obtained in \cite{ADM}.

We note that our treatment of the soft-gluon resummation for the
$\delta \phi$ distribution somewhat differs from that developed
in \cite{BN,BS}, where the resummation was performed in terms of 
a one-dimensional integral transform. We found our approach, which is 
based on a two-dimensional impact parameter ${\bm b}$ \cite{PP}, to be 
numerically 
very similar to the one of \cite{BN} near $\delta \phi = 0$, but to be 
better applicable out to larger $\delta \phi$. A more complete study of 
the Sudakov effects would start from the well-known \cite{CSS} full 
form of the Sudakov exponent, given in terms of an integral over
gluon transverse momentum. This form reduces to ours in Eq.\ (\ref{suda}) 
if the running of the strong coupling is neglected. For a 
running coupling, the Sudakov exponent becomes sensitive to
the strong-coupling regime, since $\alpha_s$ is also
probed at scales near $\Lambda_{\mathrm{QCD}}$. 
This will require the introduction of further non-perturbative 
contributions. It may also be important to determine next-to-leading 
logarithmic contributions to the exponent \cite{guffanti}. The issue 
of matching to a fixed (next-to-leading) order calculation \cite{owens} 
at larger $\delta \phi$ will then become relevant. 

We finally note that, besides the peak around $\Delta 
\phi = \pi$, the two-hadron azimuthal correlation also shows a
peak at $\Delta \phi = 0$, corresponding to the two
hadrons being in the same jet. The width of this peak
should be primarily related to the ``fragmentation part''
$\langle \hat{k}^{\perp\,2}_1\rangle/\langle z_1\rangle^2
+\langle \hat{k}^{\perp\,2}_2\rangle/\langle z_2\rangle^2$
of the width in Eq.~(\ref{rel}). One therefore expects
the peak at $\Delta \phi=0$ to be narrower than the one
at $\Delta \phi=\pi$, which indeed it is for the STAR data. 
We have checked that a distribution of the form (\ref{denom1}), 
with only fragmentation $k^{\perp}$ broadening as numerically given by 
Eqs.~(\ref{rel}) and (\ref{rel1}), fits the $\Delta \phi = 0$ peak well. 
We note that a detailed description of this region will
require using two-hadron fragmentation functions as 
studied in \cite{Sukhatme,Vendramin,ddf}.
In the following we will not consider further the same-side
peak at $\Delta \phi = 0$. 

\section{Jet correlations in $p \, p^\uparrow$ scattering}

\begin{figure}[t]
\begin{center}
\epsfig{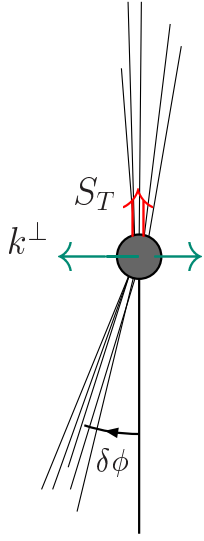}
\end{center}
\vspace*{-0.5cm}
\caption{Asymmetric jet correlation. The proton beams run
perpendicular to the drawing. \label{fig:fig2}}
\vspace*{-0.5cm}
\end{figure}

We will now study two-jet correlations near $\Delta \phi = \pi$ in
the case that one of the two incoming protons is polarized transversely 
to its momentum. As we discussed in the introduction, our
motivation is that such correlations may offer access to
the Sivers function.  The basic idea is very simple.  As follows from
Eq.~(\ref{SiversADM}), the Sivers
function represents a correlation of the form ${\bm S}_T\cdot
({\bm P} \times {\bm k}^{\perp})$ between the transverse proton 
polarization vector, its momentum, and the transverse momentum of 
the parton relative to the proton direction. In other words, if 
there is a Sivers-type correlation then there will be a preference
for partons to have a component of intrinsic transverse momentum
to one side, perpendicular to both ${\bm S}_T$ and ${\bm P}$.
Suppose now for simplicity that one observes a jet in the direction
of the proton polarization vector, as shown in Fig.~\ref{fig:fig2}.
A ``left-right'' imbalance in ${\bm k}^{\perp}$ of the parton will then 
affect the $\Delta \phi$ distribution of jets nearly opposite to the 
first jet and give the cross section an asymmetric piece around 
$\Delta\phi=\pi$. The spin asymmetry
\beq
A_{\mathrm{N}}\equiv \frac{\sigma^{\uparrow}-\sigma^{\downarrow}}
{\sigma^{\uparrow}+\sigma^{\downarrow}}
\eeq
will extract this piece. 

The denominator of this asymmetry will be, up to normalization, the 
function ${\mathcal U}$ discussed in Sec.~II. We now define 
the $y,z$ directions as given by the polarization and momentum, 
respectively, of the polarized proton. The numerator may then 
be found by considering Eq.\ (\ref{denom0}) with an additional 
factor $k_1^{\perp\, x}$ in the integrand, corresponding to the
Sivers correlation in the polarized proton:
\beq 
{\mathcal P}=\int d^2k^{\perp}_1 d^2k^{\perp}_2 \; \frac{k_1^{\perp\,
x}}{\sqrt{\langle \kappa^{\perp\,2}_1\rangle}}
\;
\delta^2 \left( {\bm k}_1^{\perp}+{\bm k}_2^{\perp} - 
{\bm P}_{j_1}^{\perp}-{\bm P}_{j_2}^{\perp}\right) \;
\bar{f}_1(k^{\perp}_1) \;f_2(k^{\perp}_2)
\; ,
\label{P0}
\eeq
where we have introduced the bar on $f_1$ to indicate that 
$\bar{f}_1$ is related to the transverse momentum dependent 
part of a Sivers function, with Gaussian width $\langle
\kappa^{\perp\,2}_1\rangle$. The latter will in general be 
different from that in the unpolarized distribution (in fact, smaller
to satisfy a positivity bound, see Ref.\ \cite{ADM}). We have
normalized $k_1^{\perp\,x}$ by $\sqrt{\langle
\kappa^{\perp\,2}_1\rangle}$ instead of by $|{\bm k}_1^\perp|$ as Eq.\
(\ref{SiversADM}) suggests. This follows the analysis of 
Ref.\ \cite{ADM} (cf.\ Eq.\ (\ref{Siversfnc}) below) and takes care of the
fact that for ${\bm k}_1^\perp=0$ the Sivers effect should vanish. 

In this way we obtain 
\beq
{\mathcal P}=\frac{r^{\perp\, x} \sqrt{\langle \kappa^{\perp\,2}_1\rangle}}
{\pi\left(\langle \kappa^{\perp\,2}_1\rangle+\langle k^{\perp\,2}_2\rangle
\right)^2} \;
\mathrm{e}^{-(r^{\perp})^2/\left(
\langle \kappa^{\perp\,2}_1\rangle+\langle k^{\perp\,2}_2\rangle\right)} \; ,
\label{P1}
\eeq
or in terms of the jet transverse momenta and azimuthal angles (measured
w.r.t.\ ${\bm S}_T$),
\beq
{\mathcal P}=\left( |{\bm P}_{j_1}^\perp| \sin \phi_{j_1} + 
|{\bm P}_{j_2}^\perp| \sin
\phi_{j_2} \right) \frac{\sqrt{\langle \kappa^{\perp\,2}_1\rangle}}
{\pi\left(\langle \kappa^{\perp\,2}_1\rangle+\langle k^{\perp\,2}_2\rangle
\right)^2} e^{-\left[ P_{j_1}^{\perp\,2}+P_{j_2}^{\perp\,2}+ 
2 P_{j_1}^{\perp} P_{j_2}^{\perp}
\cos (\Delta \phi) \right] / \left(
\langle \kappa^{\perp\,2}_1\rangle+\langle k^{\perp\,2}_2\rangle\right)}\; .
\eeq
As an example, for the case of $\phi_{j_1}=0$, corresponding to our specific
example displayed in Fig.~\ref{fig:fig2},  this yields
\beq
{\mathcal P} \propto \sin(\Delta \phi) e^{-K_{{\mathcal P}}\cos(\Delta \phi)},
\eeq
where $K_{{\mathcal P}}=2 P_{j_1}^{\perp} P_{j_2}^{\perp} /(
\langle \kappa^{\perp\,2}_1\rangle+\langle k^{\perp\,2}_2\rangle)$.

Setting $\langle \kappa^{\perp\,2} \rangle =r \langle k^{\perp\,2}\rangle$, 
the resulting spin asymmetry is
\beq \label{ansimp}
A_{\mathrm{N}} =  \frac{r^{\perp\, x}}{\sqrt{\langle k^{\perp\,2}
\rangle}} \frac{2 \sqrt{r}}{(1+r)^2}
{\mathrm e}^{-\frac{1-r}{2(1+r)}(r^{\perp})^2/\langle k^{\perp\,2}
\rangle} \; ,
\eeq
where for the denominator of this asymmetry we have taken the
function ${\mathcal U}$ given in Eq.\ (\ref{denom}) with the assumption
$\langle k^{\perp\,2}_i\rangle  
\equiv \langle k^{\perp\,2}\rangle$ for $i=1,2$. 

There are, however, several different partonic channels
contributing to jet production in $p\,p$ scattering. For
each of these, there is a combination of parton distributions,
with dependence on light-cone momentum fraction and intrinsic
transverse momentum. Thus in general, there will be a weighted
sum of $k^{\perp}$ dependent functions in the numerator and
the denominator of the asymmetry, that is,
\beq
A_{\mathrm{N}} =  \frac{\sum_{f_1,f_2} 
\int d^2 {\bm k}_1^{\perp} d^2 {\bm
k}_2^{\perp} \delta^2\left( {\bm k}_1^{\perp} + {\bm
k}_2^{\perp} - {\bm r}^{\perp} \right)\,  \frac{k_1^{\perp\,x}}{k_1^\perp}
\, \Delta^N f_1(x_1,k_1^{\perp}) \otimes
f_2(x_2,k_2^{\perp}) \otimes\,\hat{\sigma}_{f_1 f_2}(P^{\perp}_{j_1},
P^{\perp}_{j_2},\eta_{j_1},\eta_{j_2}) }
{\sum_{f_1,f_2} 
\int d^2 {\bm k}_1^{\perp} d^2 {\bm
k}_2^{\perp} \delta^2\left( {\bm k}_1^{\perp} + {\bm
k}_2^{\perp} - {\bm r}^{\perp} \right)\, f_1(x_1,k_1^{\perp}) \otimes
f_2(x_2,k_2^{\perp}) \otimes\,\hat{\sigma}_{f_1 f_2}(P^{\perp}_{j_1},
P^{\perp}_{j_2},\eta_{j_1},\eta_{j_2}) }\;.
\eeq
Here, the $\Delta^N f_1(x_1,k_1^{\perp})$ are the Sivers functions
as introduced in Eq.~(\ref{SiversADM}). The convolutions $\,\otimes\,$
are over light-cone momentum fractions only. We note that since
the Sivers functions correspond to distributions of unpolarized
partons, the hard-scattering cross sections $\hat{\sigma}_{f_1 f_2}$
are the usual unpolarized ones in both the numerator and the
denominator of the asymmetry. The pseudorapidities of the jets are denoted by 
$\eta_{j_1}$ and $\eta_{j_2}$. The hard scattering functions depend  
only on large scales, that is, on $P^{\perp}_{j_1}$ and $P^{\perp}_{j_1}$. 
Therefore, any dependence on ${\bm k}_{1,2}^{\perp}$ is neglected
in the $\hat{\sigma}_{f_1 f_2}$; in other words, one considers the 
first term in a collinear expansion. 

One can see that a simple result such as Eq.\ (\ref{ansimp})
will only emerge if the $x$ and ${\bm k}^{\perp}$ dependences
in all functions factorize from each other, if all distributions
in the numerator and, separately, in the denominator 
depend on ${\bm k}^{\perp}$ in the same way, and if the $x$-dependence
for each Sivers function is identical to that of the 
corresponding unpolarized one.

We will now give simple estimates for the possible size
of the spin asymmetry $A_{\mathrm{N}}$. For this purpose we will need
a model for the dependence of the parton distributions
on the light-cone momentum fraction $x$ as well as on
transverse momentum. We will assume that the $x$ and 
${\bm k}^{\perp}$ dependences may indeed be separated for
each function. For the unpolarized densities we write:
\beq
f (x,k^{\perp})  = f(x)\;
\frac{1}{\pi\langle k^{\perp\,2}\rangle} e^{-(k^{\perp})^2/\langle
k^{\perp\,2}\rangle} \; ,
\eeq
where the $f(x)$ are the usual unpolarized light-cone distributions
for flavors $f=u,\bar{u},\ldots,g$, for which we again use the CTEQ-5
leading order set \cite{cteq5}. For the moment, we will
neglect Sudakov effects, so we will use the value $\sqrt{
\langle k^{\perp\,2}\rangle}=2$~GeV we found in the previous section 
for the initial-state broadening without resummation of Sudakov logarithms.

The parameterizations for the Sivers function we will
use are taken from \cite{ADM}, where they were determined
from comparisons to data of Ref.\ \cite{E704} on inclusive single spin 
asymmetries for $p^{\uparrow}p\to \pi X$:
\beq
\Delta^N f (x,k^{\perp}) = 2 {\mathcal N}_f(x) \,
f(x)\,\frac{1}{\pi\langle k^{\perp\,2}\rangle}
\sqrt{2 e (1-r)} \;\frac{k^{\perp}}
{\sqrt{\langle \kappa^{\perp\,2}\rangle}}\,
e^{-(k^{\perp})^2/\langle \kappa^{\perp\,2}\rangle} \; .
\label{Siversfnc}
\eeq
We have chosen this parameterization and the accompanying fit values of 
Ref.\ \cite{ADM} since these are the only ones available so far. The question
of universality (i.e.\ process independence) of the Sivers function we leave 
as an unresolved issue. For a further discussion of this point see Sec.~IV. 

As before, $r=\langle \kappa^{\perp\,2}\rangle/\langle k^{\perp\,2}\rangle$, 
and we use $r=0.7$, in accordance with the fit of Ref.\ \cite{ADM}. This
value allows for a good fit to the E704 data \cite{E704} and also to 
$\Lambda$ polarization data, as discussed in Ref.\ \cite{ABDM}. 
The ${\mathcal N}_f(x)$ are $x$-dependent normalizations, defined 
in \cite{ADM} as
\beq \label{admfct}
{\mathcal N}_f(x) = N_f x^{a_f} (1-x)^{b_f} \frac{(a_f + b_f)^{(a_f +
b_f)}}{{a_f}^{a_f} {b_f}^{b_f}} \; .
\eeq
In Ref.\ \cite{ADM}, only the valence $u$ and $d$ Sivers functions 
were taken into account, since the data of \cite{E704} are
in the forward region of the polarized proton, corresponding to
large momentum fractions in its parton distribution functions. For 
$u$ and $d$, the parameters extracted from comparison to the 
data read \cite{ADM}:
\ba
& & N_u = 0.5, \ a_u=2.0, \ b_u =0.3 \; , \nonumber \\
& & N_d = -1.0, \ a_d = 1.5, \ b_d=0.2 \; .
\ea
For the sea quarks we assume relations identical to (\ref{admfct}), 
with ${\mathcal N}_{\bar{u}}(x)={\mathcal N}_u(x)$,
${\mathcal N}_{\bar{d},\bar{s}}(x)={\mathcal N}_d(x)$ for
simplicity. The details of these choices are not crucial. 

The size of the asymmetry is, however, very sensitive to the gluon 
Sivers function. The reason for this is that in the kinematic
regime we will investigate here, $\sqrt{s}=200$~GeV, 
$P^{\perp}_j\sim 10$~GeV, contributions from gluon-gluon and 
quark-gluon scattering are most important, because of their
large partonic cross sections and because the parton momentum
fractions become as low as $x_{1,2} \sim 0.05$. There is
so far no experimental information on the gluon 
Sivers function. It may be possible to obtain estimates 
within models of nucleon structure. This could be
an interesting topic for future studies, but is
beyond the scope of the present paper. To illustrate the
dependence of the asymmetry $A_{\mathrm{N}}$ on the
gluon Sivers function, we will simply present results for 
four distinct cases:
\begin{itemize}
\item[(i)] ${\mathcal N}_g(x)=\left( {\mathcal N}_u(x)+
{\mathcal N}_d(x) \right)/2 \;$,
\item[(ii)] ${\mathcal N}_g(x)=0\;$,
\item[(iii)] ${\mathcal N}_g(x)={\mathcal N}_d(x) \;$,
\item[(iv)] ${\mathcal N}_g(x)={\mathcal N}_d(x) \;$, but with 
$\sqrt{\langle{k^{\perp\,2}\rangle}}=2.5$~GeV for gluons.
\end{itemize}
These choices represent a very small (scenario ii), an average 
(with respect to the functions for quarks, scenario i), and two
somewhat larger (scenarios iii and iv) gluon Sivers functions.
Other choices are clearly possible, including yet larger
functions. Choice (iv) is motivated by the notion that 
intrinsic $k^{\perp}$ smearing could be larger for gluons. We keep 
$r=\langle \kappa^{\perp\,2}\rangle /\langle k^{\perp\,2}\rangle =0.7$ 
in all cases, although this choice does not result from the fit of Ref.\
\cite{ADM}, where a gluon Sivers function was not included. We have found
that a change in $r$ for the gluons of 10 \% leads to a change of about
6 \% in the asymmetry at its peak. 

\begin{figure}[t]
\begin{center}
\epsfig{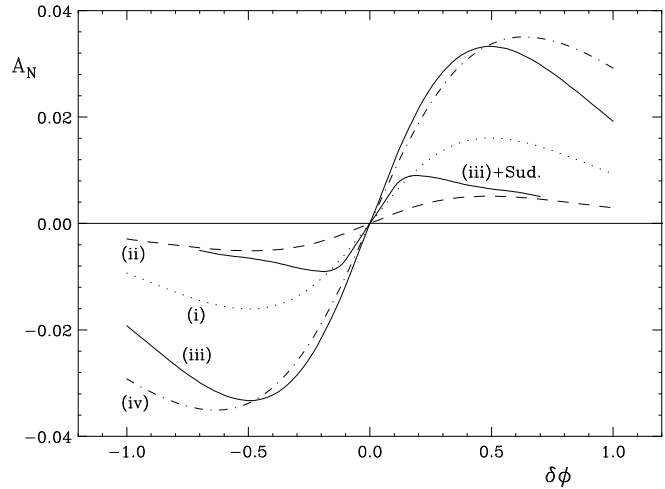}
\end{center}
\vspace*{-0.5cm}
\caption{Predictions for the spin asymmetry $A_{\mathrm{N}}$ for
back-to-back dijet production at RHIC ($\sqrt{s}=200$~GeV), 
for various different 
models (i)-(iv) for the gluon Sivers function (see text). 
The solid line marked as ``(iii)+Sud.'' shows the impact
of leading logarithmic Sudakov effects on the asymmetry
for model (iii). \label{fig:fig3}}
\vspace*{-0.5cm}
\end{figure}

The resulting asymmetries, as functions of $\delta \phi$, 
are shown in Fig.~\ref{fig:fig3}, at RHIC's $\sqrt{s}=200$~GeV. 
For simplicity, we have
chosen $\phi_{j_1}=0$. We have taken into account 
jets with pseudorapidities $|\eta_{j_{1,2}}|\leq 1$ (as suitable
for the STAR detector) and 8~GeV $\leq P^{\perp}_{j_{1,2}}
\leq 12$~GeV. The strong sensitivity to the gluon 
Sivers function is evident from Fig.~\ref{fig:fig3}.
One can see that sizable asymmetries are by all means possible. 
In fact, the asymmetry may easily be even much larger $(>10\%)$
if the gluon Sivers function is favorably close to the unpolarized
gluon density. We expect asymmetries of $\sim 1\%$ to be easily
measurable at RHIC. A typical value for the statistical uncertainty 
of such measurements may be estimated as \cite{rhicrev}
\begin{equation}
\delta A_{\mathrm{N}}\approx \frac{1}{P\sqrt{\sigma {\cal L}}} \; ,
\end{equation}
where $P$ is the beam polarization, ${\cal L}$ the integrated 
luminosity, and $\sigma$ the unpolarized dijet cross
section integrated over the kinematical bin we are interested
in.  Using $P=0.5$, a moderate ${\cal L}=10$/pb, and estimating
$\sigma=6\cdot 10^6$~pb, we find $\delta A_{\mathrm{N}}\approx
2\cdot 10^{-4}$. It will of course be important to understand
systematical uncertainties at a similar level.

It is straightforward to determine the angle where the asymmetry has its 
maximum. If we choose $\phi_{j_1}=0$ and define $\delta \phi = 
\phi_{j_2}-\pi$, then we find:
\beq
\cos(\delta \phi_{\mathrm{max}})
\approx 1 - \frac{\langle k^{\perp\,2}\rangle 
(1+r)}{2 P_{j_1}^{\perp} 
P_{j_2}^{\perp} 
(1-r)}\;,
\eeq  
which as expected is a function of the observed jet transverse 
momenta, and of $\langle \kappa^{\perp\,2}\rangle$ and 
$\langle k^{\perp\,2}\rangle$. For our parameters given above, 
this yields $\delta \phi_{\mathrm{max}} \approx 0.48$ (for scenarios i, ii 
and iii). The value of the
asymmetry at this $\delta \phi_{\mathrm{max}}$ depends of course
on the magnitude of the Sivers effect functions. 

As in the previous section we will also estimate the effect of 
Sudakov factors by including soft gluon emissions at
the leading double logarithmic level. Their effect on the denominator 
${\mathcal U}$ of the asymmetry has been described in Eq.\ (\ref{suda}). 
A difference is now that for jet pairs (unlike the case
of inclusive hadron pairs) only the initial-state
broadening plays a role \cite{greco}. This means that we will now
have $C=8/3$ for processes with only initial 
quarks and/or antiquarks, $C= 13/3$ for $qg$ scattering, and
$C=6$ for a $gg$ initial state. For
the numerator of the asymmetry one has to replace in Eq.\ (\ref{suda})
\beq
\tilde{f}_1(b^2) \to 
\int d^2 k_1^\perp \, e^{i {\bm b}\cdot {\bm k}_1^\perp} 
\frac{k_1^{\perp \, x}}{ \langle \kappa^{\perp\,2}_1\rangle}\, 
\bar{f}_1(k_1^\perp) 
= \frac{i}{2} \, b^x \, e^{-b^2 \langle \kappa^{\perp\,2}_i\rangle/4}
\equiv \frac{i}{2} \, b^x \, \tilde{\!\!\bar{f}}_1(b^2) \; .
\eeq
This leads to 
\beq
{\mathcal P} = -\frac{r^{\perp \, x}}{r^{\perp}}\,
\sqrt{\langle \kappa^{\perp\,2}\rangle} \, \int_0^\infty  
\frac{d b^2}{4 \pi} \, b J_1(b r^\perp) \, 
\exp \left[ -\frac{\alpha_s}{\pi} \,C\,\ln^2 (b \sqrt{\hat{s}}) \right]
\, \tilde{\!\!\bar{f}}_1(b^2)\tilde{f}_2(b^2) \; .
\eeq
The resulting spin asymmetry is also shown in Fig.\ \ref{fig:fig3},
for the case (iii) above, i.e., ${\mathcal N}_g(x)={\mathcal N}_d(x)$.
For this curve we have now for consistency used the smaller value 
$\sqrt{\langle k^{\perp\,2}\rangle} \approx 0.9$~GeV that we extracted 
from our Sudakov analysis of the STAR data in the previous section.
This leads to a shift of the peak of the distribution closer
to $\delta\phi=0$. More importantly, it is evident that the 
inclusion of the Sudakov factors leads to a 
considerable suppression of $A_{\mathrm{N}}$.
Of course, this does not rule out a sizable 
spin asymmetry in principle. As we mentioned earlier on, depending
on the normalization of the gluon Sivers function, we could
have a much larger $A_{\mathrm{N}}$ than given by models (i)-(iv).
In this context, we would also like to remark that the 
spin asymmetry for the single-inclusive reaction 
$p \,p^\uparrow \to \pi \, X$ at moderately 
high $p^\perp$, from which in \cite{ADM} the valence quark 
Sivers functions were extracted, is also sensitive to a small 
transverse momentum, and hence susceptible to Sudakov effects. 
The analysis of Ref.\ \cite{ADM} did not include Sudakov factors, 
which means that any effects of Sudakov suppression have effectively 
been absorbed into the Sivers function itself. In that sense 
the curve in Fig.\ \ref{fig:fig3} may include Sudakov 
suppression more than once. A Sudakov improved analysis of the 
asymmetry in $p \,p^\uparrow \to \pi \, X$ would therefore be desirable.

To gain statistics in experiment, one will not just select
events with $\phi_{j_1} \approx 0$ and vary $\phi_{j_2}$;
rather one will want to integrate over bins in $\phi_{j_1}$. Here 
one has to take care not to wash out the asymmetry by simply 
integrating over all $\phi_{j_1}$. The asymmetry will in general 
have the following dependence on $\phi_{j_1}$ and $\delta \phi$:
\beq
A_{\mathrm{N}}
(\phi_{j_1},\delta \phi) \propto \left[ |{\bm P}_{j_1}^\perp| \sin \phi_{j_1}
- |{\bm P}_{j_2}^\perp| \left(\sin \phi_{j_1} \cos \delta \phi +
\cos \phi_{j_1} \sin \delta \phi \right)\right] \, {\cal A}(\cos(\delta \phi)
).
\eeq
One possibility
is to select ``jet 1'' in the hemisphere of the ``spin-up'' direction,
and to weight the asymmetry with $\cos(\phi_{j_1})$ over this hemisphere:
\beq
A_{\mathrm{N}}
(\delta \phi) \equiv \int d \phi_{j_1}\, \cos(\phi_{j_1})\, 
A_{\mathrm{N}}
(\phi_{j_1},\delta \phi) \propto \sin \delta \phi \,
{\cal A}(\cos(\delta \phi)).
\eeq
One has to keep in mind that this weighted asymmetry has a complicated
dependence on the transverse momenta of the two jets, which differ per event. 
In general, it will be rather involved to extract the normalization and 
the width
of the Sivers functions from such a weighted asymmetry, but the above is one 
of the ways to obtain a Sivers asymmetry that is dependent on only the angle 
$\delta \phi$. More generally, one could project out the full $r^{\perp \,
x}$ azimuthal angular dependence, preferably for a fine binning in the
transverse momenta of the jets. 
 
If one were to consider leading hadrons instead of the jets, 
one would run into the problem that there could be additional effects 
generating a single spin asymmetry at leading power. In the fragmentation 
process the Collins effect could contribute \cite{Collins-93b}, which is a 
correlation between the transverse spin of a fragmenting quark and the 
transverse momentum direction of the outgoing hadron relative to that 
quark. We do not, however, think that this mechanism will be very
important here. First of all, the fragmenting quark would need
to have inherited its spin from the transverse spin of the proton.
This means that the transversity densities of the proton would be involved,
and that the partonic cross sections would depend on transverse
spin in the initial and final states. These cross sections are
much smaller than the unpolarized ones we used in our
study above \cite{sv}. In addition, a major difference between the Collins 
and Sivers effects is that the gluon Sivers function is allowed 
to be nonzero, whereas a gluonic Collins functions is forbidden
by helicity conservation. We are therefore confident that
studies of $p\, p^\uparrow \to h_1 \, h_2\, X$ (with $h_1,h_2$ two 
hadrons almost back-to-back in azimuth) at RHIC, for example in
the PHENIX experiment (where it would complement the Drell-Yan single spin
asymmetry measurements), would also be useful for learning about
the Sivers functions. 

\section{Concluding Remarks}

We have proposed an observable that could provide access to the 
Sivers effect: a single transverse-spin asymmetry in the distribution
in relative azimuthal angle $\Delta \phi \approx \pi$ of jets in a dijet 
pair. Unlike the more customary single-spin asymmetries for single-inclusive 
final states in $p\, p^{\uparrow}$ scattering, this observable is not 
power-suppressed in a large energy scale. It also has the advantage
that it will be directly (and only) sensitive to the Sivers
functions, in contrast to $A_{\mathrm{N}}$ for the process 
$p\,p^\uparrow  \to \pi \,X$ for which several different competing
mechanisms could be at work. Using experimental information on the 
average transverse momentum of initial and radiated partons from 
dimuon, diphoton, dijet and dihadron production in hadron-hadron 
collisions and using results from the study of Ref.\ \cite{ADM} on 
the Sivers effect in $p \, p^\uparrow \to \pi \, X$, we have presented 
estimates for this new observable. These indicate that the asymmetry
could well be at the few percent level, which should make it
experimentally accessible at RHIC.

Our further analysis revealed, however, that Sudakov effects
lead to a significant suppression of the asymmetry. We stress that
this does not necessarily mean that the asymmetry must be small. 
It turns out that the unknown gluon Sivers function mainly drives
the size of the asymmetry. We know of no theoretical reason why this 
distribution function should be small. In any case, any sign in 
experiment of the asymmetry we propose will be definitive
evidence for the Sivers effect. 
We also point out that the distribution in azimuthal angle 
between the jets is only one example of a variety of similar
observables in $p\,p$ scattering. Other closely related examples, 
which deserve further attention and may be equally suited for
experimental studies, are the total transverse momentum 
of the jet pair, the distribution in ``$P_{\mathrm{out}}$'' (the 
momentum of one jet out of the plane spanned by the beam axis
and the other jet's transverse momentum) \cite{BN}, and the distribution 
in ``$P^{\perp}$-balance'' $z=-{\bm P}_{j_1}^\perp\cdot 
{\bm P}_{j_2}^\perp/(P_{j_2}^{\perp})^2$ \cite{begel}. Any of these may 
be obtained from different projections in the two-dimensional 
transverse-momentum plane and may be predicted using our formulas above.

We close with a few comments on some theoretical issues, which we hope 
also provide directions for future work. As we mentioned earlier on, 
for observables that have a hard scale but additionally involve 
an observed small transverse momentum, factorization theorems are 
rather hard to establish. For our back-to-back dijet distribution, 
the issue of whether or not factorization occurs still remains
to be investigated. For the case of nontrivial polarization effects, 
such as the Sivers or Collins effects, this is a particularly 
relevant issue since, unlike in the unpolarized case, the effect 
itself already relies on the presence of non-perturbative 
``intrinsic'' transverse momentum: the parton distribution functions 
need to have an intrinsically nonperturbative dependence on the 
transverse momentum, arising as $k^\perp/M$ or $k^\perp/\langle 
k^{\perp\,2} \rangle$.

In case factorization does apply, another related complication 
is the apparent non-universality of the Sivers
functions. When it was recognized that the presence of Wilson lines
allows the Sivers functions to be non-vanishing 
\cite{Collins-02,BHS,Belitsky,BMP03}, also the remarkable result 
followed that the Sivers function in SIDIS differs by a sign 
from the one that enters in the Drell-Yan process. 
This process dependence is a unique prediction of QCD. It is
entirely calculable, but has not been studied yet for other
processes, such as $p \, p^\uparrow \to \pi \, X$ or the reaction
$p \, p^\uparrow \to 2 \, \text{jets} \, X$ we have considered here.
The color gauge invariant factorization is expected to make
definitive statements here. Jet reactions may generally be easier
to analyze theoretically than reactions with observed
hadrons in the final state, since the latter involve also
fragmentation functions that inevitably complicate
the issue of gauge invariance further \cite{BMP03,Metz}. 
A novel aspect in all this will also be the 
color gauge invariant definition of the gluon Sivers function, 
which has so far not been obtained (the same applies to any transverse 
momentum dependent gluon distribution and fragmentation functions). 
It will result in the proper gauge invariant version of 
Eq.\ (\ref{gluonDF}). For our present study, the process dependence 
implies an uncertainty. We have refrained from making any ad hoc choices 
and simply used as a starting point the valence quark Sivers functions 
obtained from $ p \, p^\uparrow \to \pi \, X$ \cite{ADM}. 

It is evident that it will be very important to deal with these 
issues. A proof of a factorization theorem for the process
$p \, p \to 2 \, \text{jets} \, X$ at small pair transverse
momentum would be highly desirable. That said, even if factorization
will be shown not to occur, the observable we have proposed is
obviously still a quantity of interest. It will presumably then give
us insight into novel aspects of QCD dynamics.

\vspace{5mm} \noindent

\begin{acknowledgments}
We are grateful to J.\ Balewski, L.\ Bland, G.\ Bunce, D.\ de Florian, 
J. Owens, S.\ Vigdor, and especially J.~Collins for stimulating 
discussions. The research of D.B.\ has been made possible by financial 
support from the Royal Netherlands Academy of Arts and Sciences.
W.V.\ is grateful to RIKEN, Brookhaven National Laboratory and the U.S.\
Department of Energy (contract number DE-AC02-98CH10886) for
providing the facilities essential for the completion of this work.

\end{acknowledgments}


\end{document}